\documentclass[12pt]{article}           

\def\preprint{UTTG-09-97}       
\def\finished{June 1997}
\def\archive {hep-th/9806059}           

\def\title{String Dualities and Toric Geometry: An Introduction  }

\long\def\abstract{
This note is supposed to be an introduction to those
concepts of toric geometry that are necessary to understand applications in 
the context of string and F-theory dualities.
The presentation is based on the definition of a toric variety in terms
of homogeneous coordinates, stressing the analogy with weighted
projective spaces.
We try to give both intuitive pictures and precise rules that should
enable the reader to work with the concepts presented here.
}

\def\CY{Calabi--Yau}
\def\ipo{\hbox{\bf 0}}

\input epsf

\usepackage{amssymb,latexsym}              \catcode`\"=\active \let"=\"

\def\ifundefined#1{\expandafter\ifx\csname#1\endcsname\relax}
\def\bye{\end{document}}   
\long\def\new#1\endnew{{\bf #1}}

\def\HS#1 {\hspace*{#1pt}} \def\VS#1 {\vspace*{#1pt}} \long\def\del#1\enddel{} 
\def\BC{\begin{center}}    
\def\EC{\end{center}}

\let\0=\over      \let\1=\vec      \def\2{{1\over2}}    \let\3=\ss
\let\4=\underline \let\5=\overline \let\6=\partial      \def\7#1{{#1}\llap{/}}
\def\8#1{{\textstyle{#1}}}         \def\9#1{{\ifmmode{\pmb{#1}}\else\bf#1\fi}}

          \def\({\left(}       \def\){\right)}

\def\eeql#1 {\label{#1}\eeq}        
\def\beq{\begin{equation}}      \def\eeq{\end{equation}}        
\def\bea{\begin{eqnarray}}      \def\eea{\end{eqnarray}}

\let\and=\wedge

\let\bra=\langle        \let\ket=\rangle        \def\<#1\>{\bra #1 \ket}
	\let\qqd=\qquad

\def\rel#1 #2{\buildrel #1 \over {#2}}  
\def\fnote#1#2{\begingroup\def\thefootnote{#1}\footnote{#2}
                \addtocounter{footnote}{-1}\endgroup}   

\def\subdef#1{\gdef\globalColor##1{##1}}      
      \subdef{Black}

\let\a=\alpha          \let\d=\delta   
           
   \let\l=\lambda  \let\m=\mu      
\let\n=\nu                  \let\r=\rho     \let\s=\sigma 
\let\t=\tau                 
               \let\S=\Sigma 
             \let\D=\Delta

\def\IR{{\mathbb R}} \def\IC{{\mathbb C}} \def\IP{{\mathbb P}} 
   
\def\IZ{{\mathbb Z}}

\def\plb#1 #2 {Phys. Lett. {\bf B#1} #2 }
\def\phr#1 #2 {Phys. Rep. {\bf  #1} #2 }        
\def\npb#1 #2 {Nucl. Phys. {\bf B#1} #2 }
\def\aph#1 #2 {Ann. Phys. {\bf #1} #2 }         
\def\jmp#1 #2 {J. Math. Phys. {\bf #1} #2 }
\def\jgp#1 #2 {J. Geom. Phys. {\bf #1} #2 }
\def\prd#1 #2 {Phys. Rev. {\bf D#1} #2 }
\def\prl#1 #2 {Phys. Rev. Lett. {\bf #1} #2 }
\def\rmp#1 #2 {Rev. Mod. Phys.  {\bf #1} #2 }
\def\zpc#1 {Z. Phys. {\bf #1C} }
\def\cmp#1 #2 {Commun. Math. Phys. {\bf #1} #2 }
\def\cqg#1 #2 {Class.Quant.Grav. {\bf #1} #2 }
\def\mpl#1 {Mod. Phys. Lett. {\bf A#1} }
\def\cpc#1 {Computer Phys. Commun. {\bf #1} }   
\def\ijmp#1 {Int. J. Mod. Phys. {\bf A#1} }
\def\ijmpC#1 {Int. J. Mod. Phys. {\bf C#1} }

\def\BP{\begin{picture}} \def\EP{\end{picture}}         
\newcounter{TRefNX} \let\OLDcite=\cite  \makeatletter
\def\makeTRefs#1{\@for  \NewTRef:=#1\do{\global\makeTRef{\NewTRef}}}
\def\makeTRef#1{\ifundefined{TRef#1}\stepcounter{TRefNX}%
\expandafter\xdef\csname TRef#1\endcsname{\theTRefNX}\fi}\makeatother
\def\NEWcite#1{\makeTRefs{#1}\OLDcite{#1}}  

\ifundefined{draftmode} {} \else        \let\cite=\NEWcite
\newcount\HOUR  \HOUR=\the\time \divide\HOUR by 60      \multiply \HOUR by 60 
\newcount\MIN   \MIN=\the\time  \advance\MIN by -\HOUR  \divide\HOUR by 60
\ifnum  \MIN>9  \def\printTIME{{\it\the\HOUR\,:\,\the\MIN}}
        \else   \def\printTIME{{\it\the\HOUR\,:\,0\the\MIN}} \fi 

\def\LLab#1{\BP(0,0)\unitlength=1mm\put(-12,.5){\makebox(0,0)[cr]{\small #1
        \rlap{$_{_{\makeatletter\csname TRef#1\endcsname\makeatother}}$}}}\EP}
\fi

\begin{document}


{\hfill \archive   \vskip -2pt \hfill\preprint }
\vskip 15mm
\centerline{\huge\bf   String Dualities and Toric Geometry:}
\vskip 8mm
\centerline{\huge\bf    An Introduction }
\begin{center} \vskip 10mm
Harald SKARKE\fnote{\#}{e-mail: skarke@zerbina.ph.utexas.edu}
\\[3mm] Theory Group, Department of Physics, University of Texas\\
        Austin, TX 78712, USA
        
\vfill                  {\bf ABSTRACT } \end{center}    \abstract

\vfill \noindent \preprint\\[5pt] \finished \vspace*{9mm}
\thispagestyle{empty} \newpage
\pagestyle{plain} 

\newpage
\setcounter{page}{1}

\section{\large Introduction}

Toric methods found their way into string theory in the winter of
92/93, when Batyrev introduced the construction of Calabi--Yau manifolds
in terms of reflexive polyhedra \cite{Bat}, relating mirror symmetry to the 
duality of polyhedra, and Witten \cite{Wph}
and Aspinwall, Greene and Morrison \cite{AGM1} discussed the phase 
structures of string compactifications.
For a while, these two topics constituted the main applications of
toric methods in the context of string theories.
More recently, it turned out that toric geometry is also a valuable
tool for the discussion of geometric properties of manifolds that become
important in the context of string dualities, such as fibration structures 
and singularities.

In the present work, we try to motivate the use of toric geometry through
its applications in the context of string and F-theory dualities and 
proceed to give an introduction to the main concepts that are relevant
in these applications. 
We will try to give both intuitive pictures and precise rules that should
enable the reader to work with the concepts presented here.

Our presentation will revert the historical order: 
Originally, toric geometry was defined in terms of rather abstract
algebraic concepts (semigroups, ideals, Spec, \ldots) and only later
it was found by several authors (perhaps in the clearest form by Cox in
\cite{Cox1}) that toric varieties admit global homogeneous variables 
in a way that is very similar to (weighted) projective spaces.
This construction is also the one that is used for discussions of phase
structures \cite{Wph, AGM1}.
As noted by Cox in \cite{Cox2}, ``It is possible to develop the
entire theory of toric varieties using \ldots as the {\it definition} of
toric variety''.
This is the path we will follow (without too much mathematical
rigour, however), and it is surprising how many statements can be derived
or at least explained without using a large apparatus of algebraic geometry.
Important sources of further information on toric geometry are the textbooks
by Fulton \cite{Ful} and Oda \cite{Oda} and the recent review by Cox 
\cite{Cox2}; introductions to toric
geometry intended for physicists can be found in \cite{AGM2,HKT,MP1,Gre}.
Inevitably, the presentation will be influenced mainly by the `Austin style'
\cite{CF}--\cite{CPR4}; other applications of toric geometry to string 
and F-theory dualities can be found in refs. \cite{KM}--\cite{BKMT}.

We will start with giving a brief overview of F-theory and string dualities
in the next section, mainly with the aim of explaining which geometric
structures play a role in string dualities.
In section 3 we introduce toric varieties as generalisations of
weighted projective spaces, and in section 4 we discuss singularities
and construct coordinate patches.
We proceed to explain how to construct functions and line bundles on
toric varieties in section 5 and how fibrations structures manifest themselves
torically in section 6. 
Finally, in section 7 we return to the subject of singularities and explain 
how enhanced gauge groups in type IIA and F-theory can be read off from the 
toric polyhedra.

\section{\large F-theory and string dualities}

In this section we give a brief account of Vafa's construction of F-theory
\cite{F} as a particular eight dimensional vacuum of the type IIB string.

Type IIB string theory in ten dimensions is chiral with 
two left-moving and no right moving space-time supersymmetries.
The bosonic fields of the corresponding low energy field theory are 
the graviton $g_{\m\n}$, the antisymmetric tensor field $B_{\m\n}$ and 
the dilaton $\phi$ coming from the NS-NS sector of the string theory, as well
as the axion $\tilde \phi$ and the antisymmetric tensor fields 
$\tilde B_{\m\n}$ and $A^+_{\m\n\r\s}$ (the latter being self-dual)
coming from the RR sector.
This theory has a well known conjectured non-perturbative $SL(2,\IZ)$
symmetry \cite{HT,Sch1}.
Under this symmetry, the combination
$ \t=\tilde \phi + i e^{-\phi} $ 
of the axion and the dilaton, and the doublet $(B,\tilde B)$ of two form fields
are believed to transform as
\beq \t\to{a\t+b\0 c\t+d},\qqd 
     {B\choose \tilde B} \to {a\;b\choose c\;d}{B\choose \tilde B}
\eeql{trafo}
respectively, while $g$ and $A^+$ remain invariant.

Vafa looked for a solution of the low energy field equations such that
$B$, $\tilde B$ and $A^+$ vanish but $G$ and $\t$ depend on the space-time
coordinates $x_8$ and $x_9$, but not on $x_0,\ldots,x_7$.
Demanding that such a solution should also preserve half of the supersymmetry
results in a BPS condition implying that $\t$ depend holomorphically on 
$z=x_8+ix_9$. 
The low energy lagrangian for $\t=\t(z)$ 
\del
turns out to be 
\beq S=\int\sqrt{g}\(R+{\6\t\6\bar\t\0({\rm Im}\t)^2}\). \eeq
This lagrangian 
\enddel
allows a solution where $z$ parametrizes a Riemann sphere
$\IP^1$ (complex projective one dimensional space, often denoted $CP^1$)
and $\t(z)$ has generically 24 singularities where
$\t(z)\sim {1\0 2\pi i} \ln(z)$.   
We see immediately that $\t$ is not, strictly speaking, a function of $z$,
since 
\beq \t\to\t+1\qqd\hbox{ for }\qqd z\to e^{2\pi i} z.  \eeql{deg}
Nevertheless this is a good vacuum for type IIB string theory once we take 
into account the action (\ref{trafo}) of the $SL(2,\IZ)$ symmetry.

A quantity $\t$ that is defined only up to transformations of the
type (\ref{trafo}) has a well known geometric interpretation: 
It can be seen as the complex structure modulus of a torus $T^2$.
Thus, Vafa's solution of the IIB string theory corresponds to a $\IP^1$
with a two-torus $T^2$ `on top of every point'.
The total space parametrized locally by $(z,w)$,
where $w$ denotes a coordinate on the $T^2$, has to be a $K3$ surface.
The complex structure of the $T^2$ varies over the $\IP^1$ and degenerates
over 24 points in the manner dictated by eq. (\ref{deg}).
If some of the 24 points collide, we get singularities that  are worse
than the one described by (\ref{deg}).
Physically each of the 24 points corresponds to the location of a 7-brane.

A structure of the type described above is not uncommon in algebraic geometry
and is known under the name `fibration'.
Generally a fibration manifests itself through a surjective map from the 
total fibration space to the base space such that the preimage of a generic 
point in the base is a copy of the fiber, with the moduli of the fiber 
depending holomorphically on the coordinates of the base.

The resulting eight dimensional theory may be seen as the compactification of 
a 12 dimensional theory on the K3 surface parametrized by the 
2 complex coordinates $z$ and $w$.
This 12 dimensional theory is known as F-theory.

As a theory in eight dimensions, it has exactly half of the maximally possible
supersymmetry.
There is another theory in eight dimensions with the same amount of unbroken 
supersymmetry, namely heterotic string theory compactified on $T^2$, so it
is natural to conjecture a duality.
Indeed, there are very good reasons to believe that there exists such a 
duality, with the heterotic coupling constant given by the size of the $\IP^1$
\cite{F}.

This theory can now be compactified further to 6 dimensions.
There are two distinct possibilities: If we do not want to break
supersymmetry further, we may compactify on a further $T^2$.
Then the aforementioned duality becomes a duality between F-theory on 
$K3\times T^2$ and heterotic string theory on $T^2\times T^2$.
The latter theory is well known to appear in another duality relating it
to type IIA string theory on $K3$ \cite{HT,Wstd}.
It is believed that the two $K3$ manifolds occurring in these dualities
are the same. 
This is very useful since the mechanism for the occurrence of enhanced
gauge groups in type IIA is well understood.

Alternatively, we may compactify the 8 dimensional theory on a $\IP^1$
in such a way that the complex structure of the K3 depends holomorphically
on the complex coordinate of the new $\IP^1$.
In this case we compactify F-theory on a Calabi-Yau threefold which is
both K3 fibered and elliptically fibered.
This theory is dual to heterotic string theory compactified on a K3.
By compactifying this theory to four dimensions on a $T^2$, we may again 
use a known duality \cite{KV} to relate it to a type IIA string theory.

\section{\large Toric varieties as generalisations of weighted projective 
spaces}

Algebraic geometers think of two-tori as `elliptic curves'.
A standard way of describing an elliptic curve is by embedding it
into 
\beq \IP^2=(\IC^3\setminus \{0\})/(\IC\setminus \{0\}),\eeq
where the division by $\IC\setminus \{0\}$ means that we identify points 
related by the equivalence relation $\sim$ acting in $\IC^3\setminus \{0\}$ 
through
\beq (x,y,z)\sim (\l x,\l y, \l z)\;\hbox{ for any }\; 
      \l \in  \IC\setminus \{0\}; \eeql{eqp2}
$x$, $y$ and $z$ are called homogeneous coordinates.
The elliptic curve is embedded in this space via the Weierstrass 
equation 
\beq y^2z=x^3+axz^2+bz^3. \eeq
An alternative description can be given in terms of the weighted 
projective space $\IP^{(2,3,1)}$ defined just like $\IP^2$, but with the
equivalence relation changed to 
\beq (x,y,z)\sim (\l^2 x,\l^3 y, \l z)\;\hbox{ for any }\; 
      \l \in  \IC\setminus \{0\} \eeql{eqwp2}
and the Weierstrass equation changed to
\beq y^2=x^3+axz^4+bz^6. \eeq
In both cases this equation describes a non-singular elliptic curve
whenever the discriminant $\d:=4a^3+27b^2$ is nonvanishing.

\begin{figure}[htb]
\epsfxsize=2.5in
\hfil\epsfbox{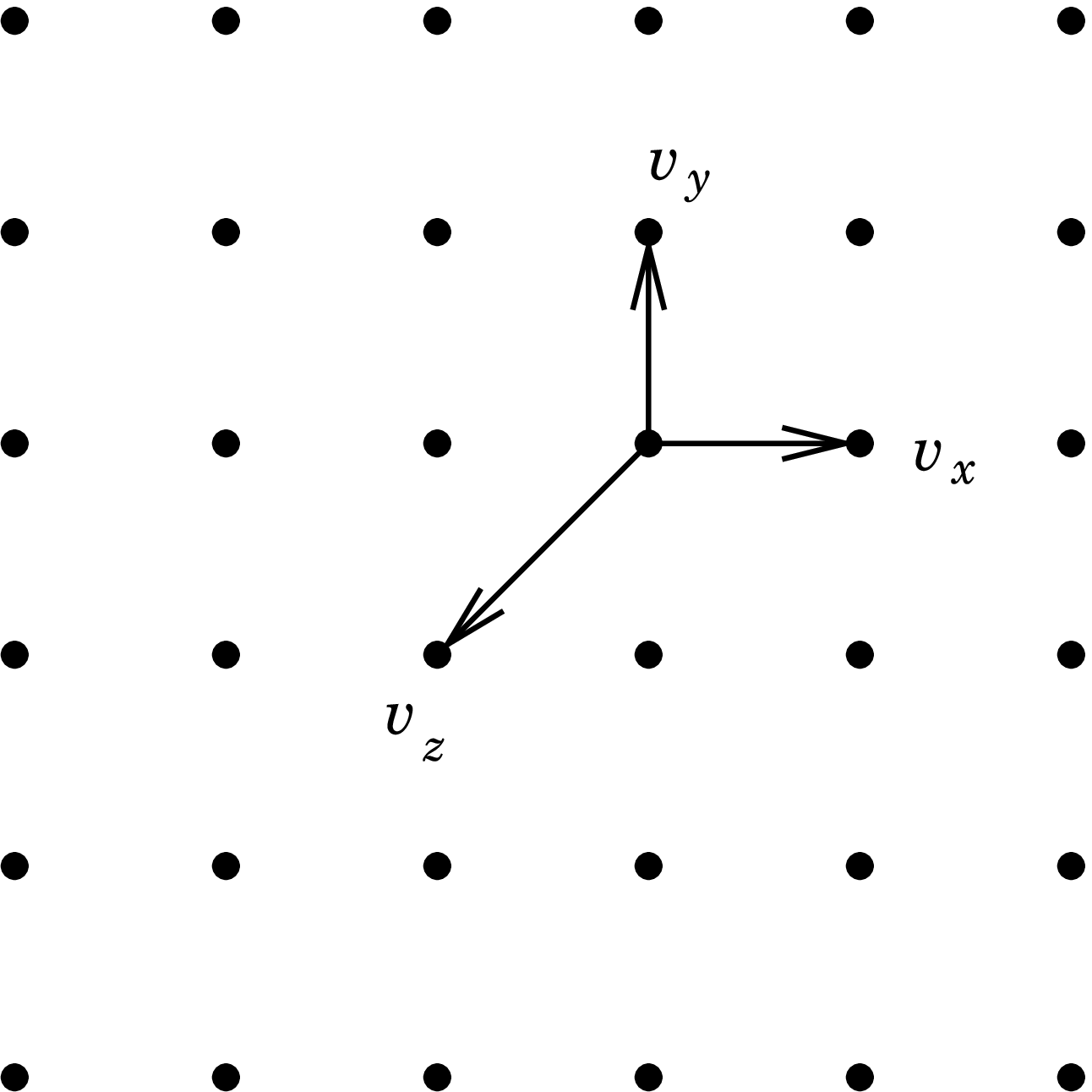}\hfil
\epsfxsize=2.5in
\hfil\epsfbox{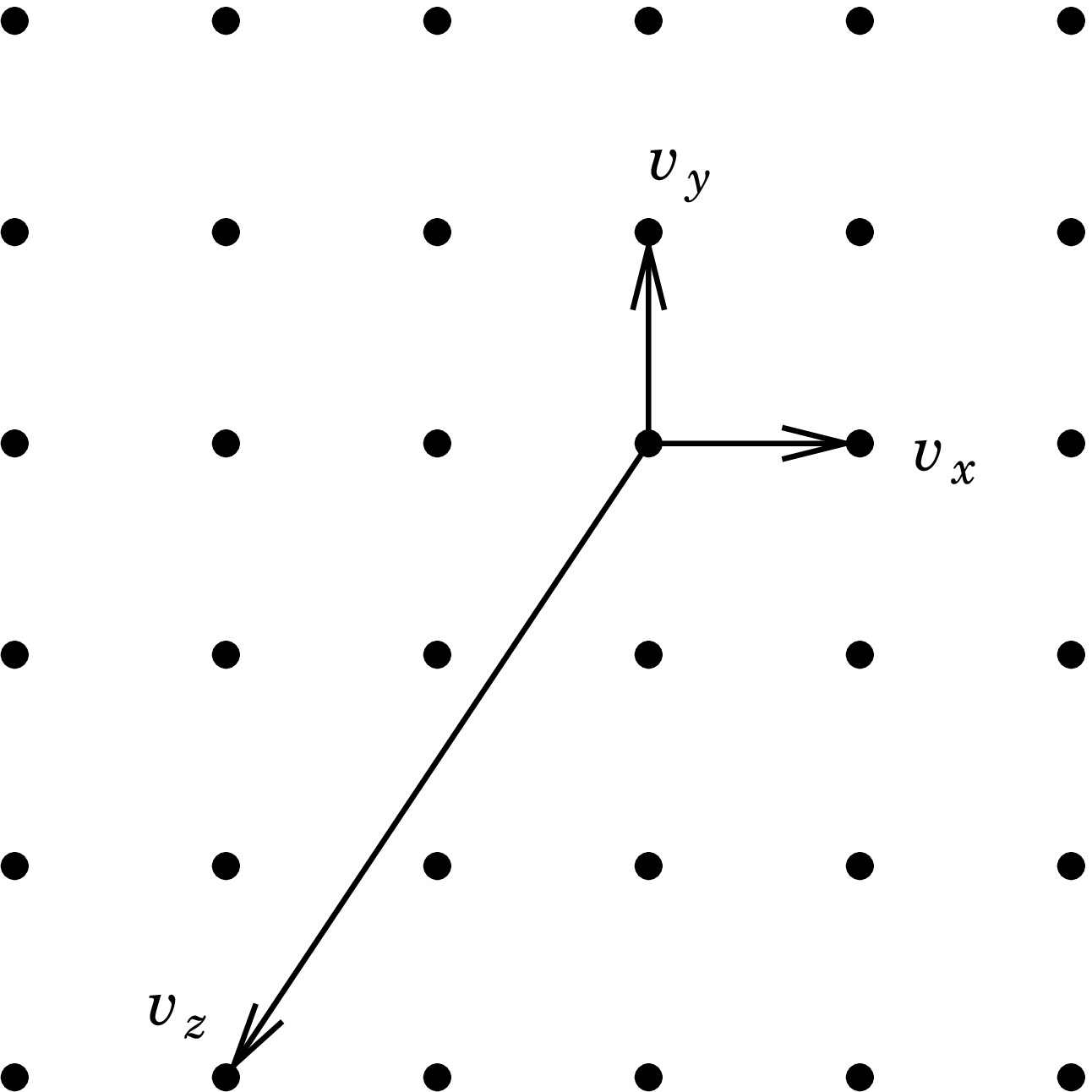}\hfil
\caption{The fans for $\IP^2$ and $\IP^{(2,3,1)}$ }
\label{fig:fans}
\end{figure}

$\IP^2$ and $\IP^{(2,3,1)}$ are simple examples of toric varieties.
The equivalence relations (\ref{eqp2}) and (\ref{eqwp2}) can be 
encoded in diagrams like those in fig. \ref{fig:fans}.
In both diagrams we have drawn vectors $v_x$, $v_y$ and $v_z$
in some lattice called the `$N$ lattice', such that 
\beq q_xv_x+q_yv_y+q_zv_z=0,\eeql{ws}
where $q_x$, $q_y$ and $q_z$ are just the exponents of $\l$ in (\ref{eqp2}) 
and (\ref{eqwp2}), respectively.
It might be helpful to think about this in the following way: For
$\IP^{(2,3,1)}$, for example, the fact that multiplying $x$ by $\l^2$,
$y$ by $\l^3$ and $z$ by $\l$ takes us back to the same point in 
$\IP^{(2,3,1)}$ is encoded in the toric diagram by the fact that adding 
$2v_x$, $3v_y$ and 
$v_z$ to a lattice point takes us back to the same lattice point.
So far this might look like a fancy but useless way of memorising
the structure of a weighted projective space, but we will see how
powerful this construction is in a moment.

More generally, we have the following situation: 
There are $k\ge n$ vectors $v_i$ in a lattice $N$ isomorphic to $\IZ^n$, 
and $k-n$ independent linear relations of the type (\ref{ws}) leading to
equivalence relations like (\ref{eqp2}, \ref{eqwp2}).
Thus the complex dimension of the toric variety is always equal to the real 
dimension of the lattice.

In more than two dimensions the vectors $v_i$ are not sufficient to
determine the structure of a toric variety, and we need a few definitions: 
We define a strongly convex rational polyhedral cone to be an
$n$ or lower dimensional cone in $N_\IR$ (the real vector space carrying 
the lattice $N$), with $\ipo\in N_\IR$ as its apex,  with the following 
properties:
it is bounded by finitely many hyperplanes (`polyhedral'), 
its edges are spanned by lattice vectors (`rational'),
and it contains no complete line (`strongly convex').
A face of a cone $\s$ is either $\s$ itself or the intersection of $\s$
with some hyperplane bounding $\s$.
A fan $\S$ is defined to be a collection of cones such that 
with every cone it contains it also contains any face of it and 
that the intersection of any two cones in $\S$ is a face of each
($\ipo$ is also considered to be a cone).
In each of our examples there is the zero dimensional cone consisting of 
the origin only, there are three one dimensional cones given by the
rays determined by $v_x$, $v_y$ and $v_z$, and there are the two dimensional 
cones corresponding to the segments into which these rays cut 
$N_\IR\simeq \IR^2$.

The generalisation of our previous construction of (weighted) $\IP^2$
to arbitrary toric varieties is as follows:
To each one-dimensional cone in $\S$ with primitive generator $v_k$ we 
assign a homogeneous coordinate \cite{Cox1} $z_i$, $i=1,\cdots,k$.
{}From the resulting $\IC^k$ we remove the exceptional set
\beq Z_\S=\bigcup_I\{(z_1,\cdots,z_k):\; z_i=0\;\forall i\in I\}\eeq
where the union $\bigcup_I$ is taken over all sets 
$I\subseteq \{1,\cdots,k\}$ for which 
$\{v_i:\;i\in I\}$ does not belong to a cone in $\S$.
This can be rephrased
as the statement that several $z_i$ are allowed to vanish 
simultaneously only if the corresponding $v_i$ belong to the same cone.
Then our toric variety 
is given by the quotient of 
$\IC^k\setminus Z_\S$ by a group which is the product of a finite abelian 
group and $(\IC\setminus \{0\})^{k-n}$ acting by 
\beq (z_1,\cdots,z_k)\sim (\l^{q^1_j}z_1,\cdots,\l^{q^k_j}z_k)~~~~~
{\rm if}~~~~\sum_{i=1}^k q^i_j v_i=0 \eeql{er}
($k-n$ of these linear relations are independent; the $q^i_j$ are chosen 
such that they are integer and the greatest common divisor of the $q^i_j$ 
with fixed $j$ is 1).
This definition will become much clearer after we have discussed 
singularities and coordinate patches in the next section.
Then we will also state what the finite abelian group (which is trivial in 
most examples, anyway) is.

\section{\large Singularities, blow-ups and coordinate patches}

Toric varieties often have singularities.
As an example let us again consider $\IP^{(2,3,1)}$.
Near the point $y=z=0$ we may use the equivalence relation (\ref{eqwp2})
to set $x=1$.
This does not use up all of our freedom in choosing $\l$, since we are
still left with the freedom of choosing a sign for $\l$.
Thus we are left with a residual relation $(y,z)\sim(-y,-z)$, and
our toric variety looks locally like $\IC^2/\IZ_2$.
In algebraic geometry there exists a procedure for turning a singular 
variety into a regular one, known as blow-up: 
A point (or, more generally, a subvariety) is said to be blown up if it is
replaced by a higher-dimensional subvariety.

\begin{figure}[htb]
\epsfxsize=2.5in
\hfil\epsfbox{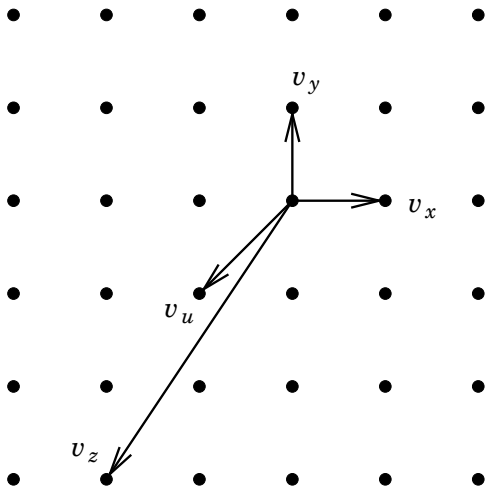}\hfil
\caption{The fan for the blow-up of $\IP^{(2,3,1)}$}
\label{fig:bu}
\end{figure}

In our present example, this is very easily visualised torically:
Consider replacing the fan for $\IP^{(2,3,1)}$ as in fig. \ref{fig:fans}
by the fan in fig. \ref{fig:bu}.
According to the rules of the previous section, our toric variety is given by
\beq \IC^4\setminus\{(x,y,z,u):\;(x=u=0)\lor(y=z=0)\} \eeql{xyzu}
divided by the equivalence relation
\beq (x,y,z,u)\sim (\l^2\m x,\l^3\m y, \l z, \m u)\;\hbox{ for any }\; 
      (\l,\m) \in  (\IC\setminus \{0\})^2. \eeql{eqbu}
This variety can be analysed in the following way:
Wherever $u$ is not equal to zero, we can use $\m$ to set $u=1$,
thus obtaining a copy of $\IP^{(2,3,1)}$ with the singular point $y=z=0$
removed.
At $u=0$, we have $x\ne 0$ according to (\ref{xyzu}).
If we switch from $(\l, \m)$ to $(\l, \r)$ with $\r=\l^2\m$ and use our
freedom in choosing $\r$ to set $x$ to 1, we are left with a space described by
\beq (1,y,z,0)\sim (1,\l y, \l z, 0). \eeq
This is just a $\IP^1$ parametrized by $y$ and $z$ and is called an
`exceptional divisor'.
In general a divisor in an algebraic variety is a formal linear combination 
of irreducible hypersurfaces with integer coefficients.
In the toric case we always have the divisors corresponding to the 
hypersurfaces obtained by setting one of the homogeneous coordinates to zero.
In this way we may think of the vectors $v_i$ as corresponding to the 
`toric divisors' determined by $z_i=0$.
Similarly, higher dimensional cones correspond to lower dimensional 
algebraic subvarieties.

The reader is invited to check that there is a singularity at $x=z=0$
looking locally like $\IC^2/\IZ_3$ that can be blown up by introducing two 
extra divisors corresponding to rays between $v_x$ and $v_z$.

We may think of $n$-dimensional cones as representing coordinate 
patches and of lower dimensional cones as representing the regions of 
overlap along which these patches are glued, in the following way:
Consider a simplicial $n$ dimensional cone, i.e. a cone 
generated by $n$ independent vectors which we choose to call $v_1,\ldots,v_n$.
To this cone there corresponds in a natural way the part of the toric variety
where the $z_i$ with $i>n$ are non-zero, but some or all of $z_1,\ldots,z_n$
may be zero.
We may choose the $q^i_j$ of (\ref{er}) in such a way that they correspond to
expressing each of the $v_i$ with $i>n$ in terms of $v_1,\ldots,v_n$.
Explicitly, this means 
\beq q_j^{n+j}v_{n+j}+\sum_{i=1}^nq_j^iv_i=0\;\hbox{ for }\;j=1,\ldots,k-n.\eeq
If $v_1,\ldots,v_n$ generate the lattice $N$, then the $q_j^{n+j}$ can always
be chosen to be 1 and the corresponding $\l_j$ may be used to set $z_{n+j}$ 
to 1 without any further freedom remaining.
If, however, $v_1,\ldots,v_n$ generate only a sublattice $M(v_1,\ldots,v_n)$
of $M$, residual relations generalising the ones in our example may occur.
These relations belong to a finite abelian group $G(v_1,\ldots,v_n)$ which is
isomorphic to $M/M(v_1,\ldots,v_n)$. 
If all $v_i$ generate $M$, then our coordinate patch will look like 
$\IC^n/G(v_1,\ldots,v_n)$.
In the case where $M$ is not generated by all of the $v_i$, we define the
toric variety in such a way that this result still holds:
Then we have to divide $\IC^k\setminus Z_\S$ not only by $(\IC-\{0\})^{k-n}$
as in (\ref{er}) but also by the finite group $G(v_1,\ldots,v_k)$.

Thus we have seen that whenever $\S$ is simplicial, the corresponding variety 
will have only orbifold singularities, and in particular that it is smooth 
if every $n$ dimensional cone is generated by vectors that generate the 
lattice $M$.
The case of cones that are not simplicial is more complicated and will not be
considered here. It is always possible to subdivide a fan to make it 
simplicial, anyway.

\section{\large Functions and line bundles}

We now turn to the subject of functions on toric varieties.
To this end, consider the lattice $M$ dual to our original lattice $N$.
A vector $w\in M$ may be used to define a Laurent monomial 
\beq x^{\<v_x,w\>} y^{\<v_y,w\>} z^{\<v_z,w\>}, \eeql{laur}
where $\<v,w\>$ denotes the duality pairing between $v\in N_\IR$ and
$w\in M_\IR$.
To avoid confusion it might be helpful to remember that $M$ means Monomial
and $N$ (somewhat less naturally) means faN.
(\ref{laur}) describes the case of a weighted $\IP^2$; the generalisation to 
an arbitrary toric variety is obvious.
Under $x\to \l^{q_x}x,\ldots$, this monomial becomes
\beq (\l^{q_x}x)^{\<v_x,w\>} (\l^{q_y}y)^{\<v_y,w\>} (\l^{q_z}z)^{\<v_z,w\>}
   = \l^{\<q_xv_x+q_yv_y+q_zv_z,w\>}x^{\<v_x,w\>} y^{\<v_y,w\>} z^{\<v_z,w\>}.
\eeq
As $q_xv_x+q_yv_y+q_zv_z=0$, (\ref{laur}) is invariant under the equivalence 
relation and therefore is a true meromorphic function on our variety.

\begin{figure}[htb]
\epsfxsize=2.5in
\hfil\epsfbox{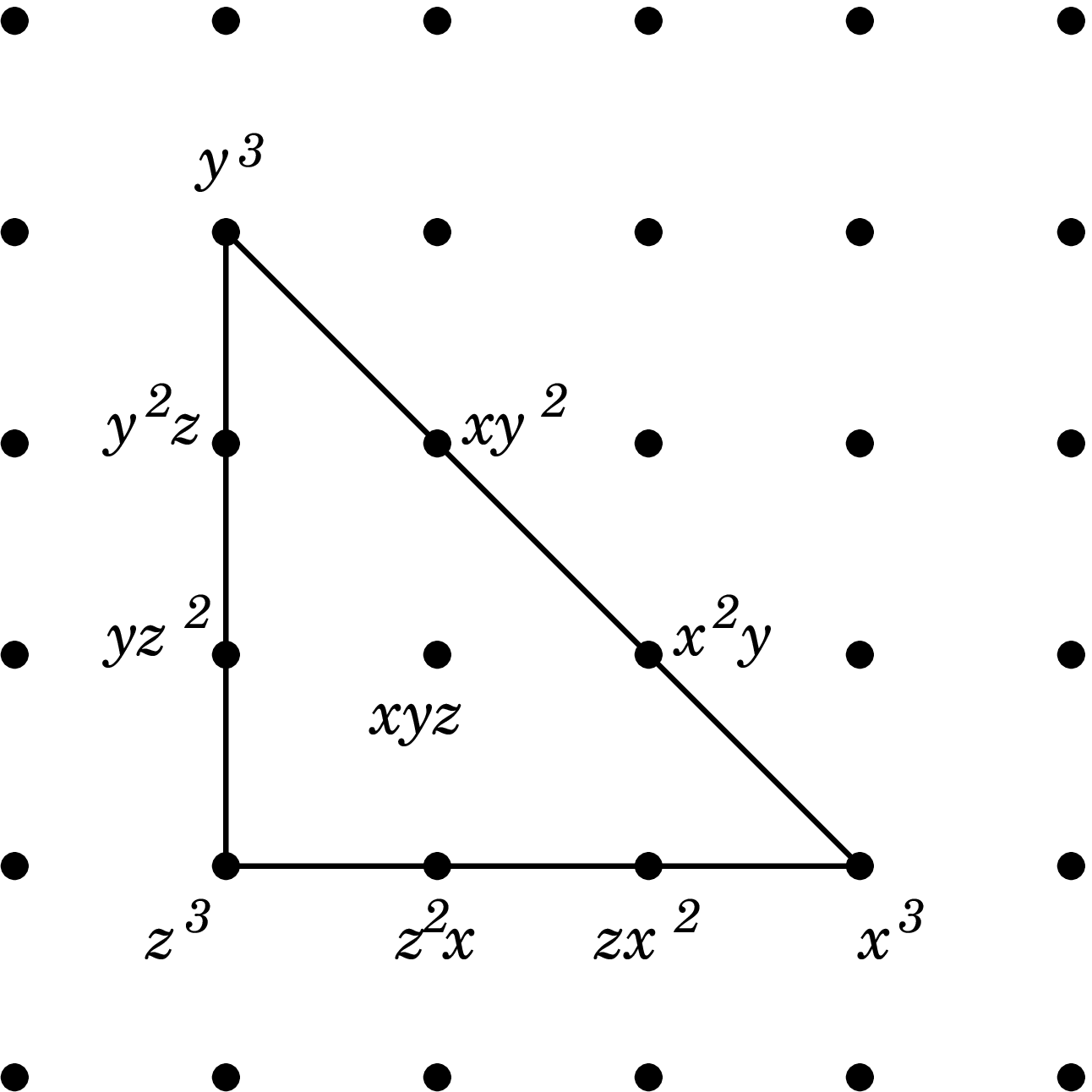}\hfil
\epsfxsize=2.5in
\hfil\epsfbox{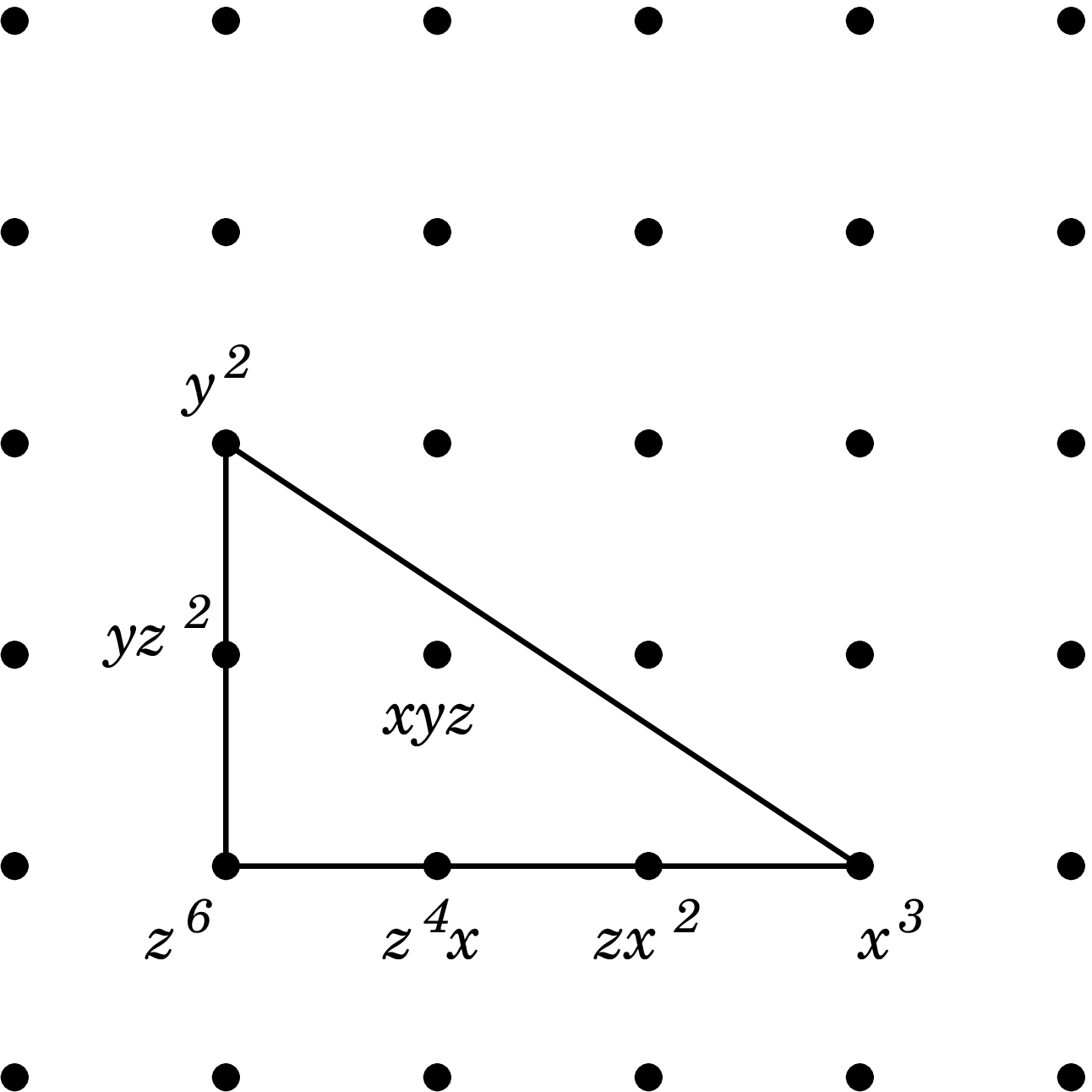}\hfil
\caption{The polyhedra corresponding to $P$ for $\IP^2$ and $\IP^{(2,3,1)}$ }
\label{fig:m}
\end{figure}

In order to define a consistent equation on a toric variety, however, we do 
not need functions:
A polynomial equation $P(z_i)=0$ is well defined if $P$ transforms as
$P\to \l^{q_p}P$ under $z_i\to\l^{q_i}z_i$.
For example, 
\beq P(x,y,z)=\sum_wa_wx^{\<v_x,w\>+1} y^{\<v_y,w\>+1} z^{\<v_z,w\>+1} 
\eeql{clb}
transforms homogeneously with $q_P=q_x+q_y+q_z$.
Once again, the generalisations (arbitrary toric varieties, replacing the 1's
by other numbers) are obvious.
$P$ is holomorphic (i.e., has no poles) if all $w$ for which the coefficient 
$a_w$ is non-vanishing
fulfill
\beq \<v_i,w\>\ge -1 \;\hbox{ for all }i.\eeql{vw}
These inequalities define a bounded region (a polygon or, more generally,
a polyhedron) in the real vector space $M_\IR$ carrying the lattice $M$.
Figure \ref{fig:m} shows these regions for our two favourite examples,
with points labeled by the monomials they represent according to (\ref{clb}).


We may now ask the following question: Given a specific polynomial $P$ in
the homogeneous coordinates of some toric variety, which toric blow-ups are 
consistent with it?
The answer is again given by the inequality (\ref{vw}), this time seen
as a condition on the allowed $v_i$, given the $w$ that encode $P$.
Denoting by $\D$ the polyhedron in $M_\IR$ that is the convex hull of
the lattice points corresponding to some monomials, we may define the
dual polyhedron to be
\beq \D^*:=\{v\in N_\IR:\;\<v,w\>\ge -1 \;\hbox{ for all }w\in \D\}. \eeq
Then $\D$ is called reflexive if $\D^*$ is a lattice polyhedron (i.e.
if the vertices of $\D^*\subset N_\IR$ lie in $N$).
In our examples, $\D^*$ is just the convex hull of the endpoints of the
vectors $v_x$, $v_y$ and $v_z$ (cf. fig. \ref{fig:fans}), and the allowed 
blow-ups are the one we encountered already and two more along the line 
$v_xv_z$ in the second picture of fig. \ref{fig:fans}.

Let us now consider the toric variety determined by a fan containing all
rays determined by points in $\D^*$ and let us also assume that this
fan is maximally triangulated (i.e., no  cone can be 
subdivided into smaller cones without introducing extra rays).
It was shown by Batyrev \cite{Bat} that the hypersurface defined by 
the vanishing of a generic polynomial in the class determined by $\D$
is a smooth Calabi-Yau manifold for $n\le 4$.
For $n\le 3$ reflexivity ensures smoothness of the underlying toric variety, 
whereas for $n=4$ (Calabi-Yau threefolds) the toric variety may have 
point-like singularities which are however missed by the generic hypersurface.

The following paragraph, which will not be necessary for understanding
the remaining two sections, requires a little knowledge about line bundles
and their connections with divisor classes, at the level of e.g. chapter
1.1 of \cite{GH}.
It is easily checked that (\ref{clb}) defines a section of a line 
bundle, with other sections being determined by different values of the $a_w$.
The corresponding divisor class can be read off from any of the monomials in
$P$; for the particular form (\ref{clb}) we always have $xyz$ among the 
monomials and so the divisor class is $[D_x+D_y+D_z]$, with $D_x=\{x=0\}$ etc.
Moreover, for any toric variety $[\sum_i D_i]$ (the sum ranging over all 
toric divisors $z_i=0$) is equal to the anticanonical class of the variety 
(this can be shown \cite{Ful} with methods very similar to those for 
determining the anticanonical class of $\IP^n$).
Thus a polynomial $P=\sum_w\a_w\prod_iz_i^{\<v_i,w\>+1}$ will always determine
a section of the anticanonical bundle of the toric variety, and by the 
adjunction formula the zero locus of such a section describes a variety of 
trivial canonical class.
If this variety is smooth, which is guaranteed by reflexivity for $n\le 4$,
it is thus a Calabi-Yau manifold.

\section{\large Fibrations}

\begin{figure}[htb]
\epsfxsize=3in
\hfil\epsfbox{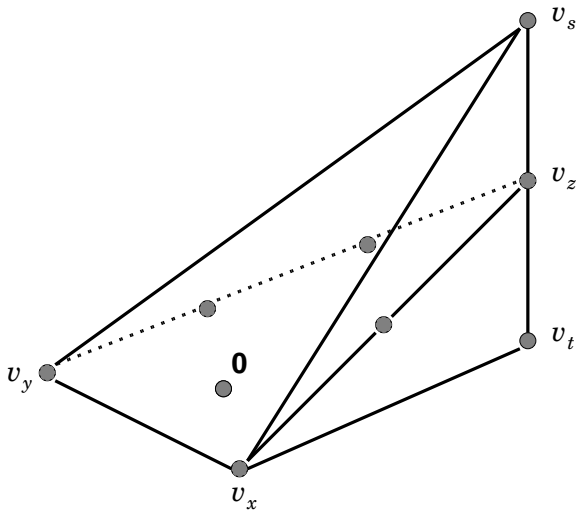}\hfil
\caption{The polyhedron $\D^*$ corresponding to a smooth elliptically fibered
K3 surface}
\label{fig:wei}
\end{figure}

In this section we consider the question of how fibration structures can
be described torically.
We remember that a fibration has a base space and a fiber, with the complex
structure of the fiber depending on the point of the base.
The simple example of an elliptic fibration with base $\IP^1$ can be
described by a Weierstrass equation for the fiber, with coefficients 
depending on homogeneous coordinates $(s,t)$ for the base.
Such an equation may take the following form:
\beq y^2=x^3+a(s,t)xz^4+b(s,t)z^6. \eeql{fibr}
What should the corresponding toric diagrams look like?
They should be three--dimensional, since we want to embed a complex surface
with a single equation; and as $x$, $y$ and $z$ must still satisfy the
equivalence relation there must again be vectors $v_x$, $v_y$ and $v_z$
with $2v_x+3v_y+v_z=0$.
In addition we need two more vectors representing the new
coordinates $s$ and $t$.
Fig. \ref{fig:wei} shows a polytope $\D^*$ such that the fan $\S$ over 
a triangulation of $\D^*$ has these properties.
There is the additional linear relation $v_s+v_t=2v_z$, so that the toric 
variety in which the Weierstrass equation is defined is given by quintupels 
$(x,y,z,s,t)$ subject to equivalence relations
\beq (x,y,z,s,t)\sim (\l^2 x,\l^3 y, \l\m^{-2} z, \m s, \m t)\;
     \hbox{ for any }\;     (\l,\m) \in  (\IC\setminus \{0\})^2. \eeql{eqk3}
By constructing the dual polyhedron $(\D^*)^*=:\D\subset M_\IR$ we can find out
which monomials are allowed to occur.
The result is 
\beq a(s,t)=a_0s^8+a_1s^7t+\ldots+a_8t^8,\qqd
     b(s,t)=b_0s^{12}+b_1s^{11}t+\ldots+b_{12}t^{12},  \eeq
so the discriminant $\d=4a^3+27b^2$ is of degree 24 and is thus expected to
vanish at 24 points $(s,t)\in \IP^1$.

In general, fibration structures involving manifolds of vanishing first 
Chern class both as fiber and as total space manifest themselves by
$\D^*_{\rm fiber}$ being a reflexive subpolyhedron of $\D^*_{\rm total}$.
The fan of the base space may be determined by projecting the fan of the 
total space along the directions of $\D^*_{\rm fiber}$ \cite{fft}.

\section{\large Singularities revisited}

As we saw in the previous section, if we choose the polynomials $a$ and $b$
to be homogeneous of orders 8 and 12, respectively, the
fibration will generically degenerate over 24 points of the base $\IP^1$.
At these points the elliptic fiber degenerates, but the K3 surface is
still smooth.

If we allow $a$ and $b$ to take special forms, then the singularity type 
of the fiber may get worse and the K3 surface can become singular, too.
There exists a classification of degenerations of elliptic fibers 
according to  Kodaira.
In this scheme the fiber type is determined by the orders of vanishing 
of $a$, $b$ and $\d$.
The generic case corresponds to 24 singularities where 
$(o(a),o(b),o(\d))=(0,0,1)$,
called $I_1$ singularities, whereas every other fiber is of type $I_0$,
which is nothing but a smooth fiber.
Generally $I_n$ fibers are defined by $(o(a),o(b),o(\d))=(0,0,n)$.
If we restrict $a$ and $b$ to be
\beq a(s,t)=a_4s^4t^4,\qqd
     b(s,t)=b_5s^5t^7+b_6s^6t^6+b_7s^5t^7,  \eeql{e8e8}
one can easily check that $(o(a),o(b),o(\d))=(4,5,10)$ both at $s=0$
and $t=0$.
In this case we have two $II^*$ singularities.

Alternatively, we may discuss singularities of the K3 surface independently of 
the fibration structure. 
In toric terms, we may think about singularities in the following way:
Restricting $a$ and $b$ means that the polyhedron $\D$ in the $M$ lattice
becomes smaller.
This implies that we can enlarge $\D^*$, thereby introducing exceptional 
divisors.
The intersection matrix of these exceptional divisors turns out to be 
the Cartan matrix of a simple laced, i.e. $A$, $D$ or $E$ Lie algebra.
In this way we have made the connection with the ADE classification of 
singularities of surfaces.
By some well known results on compactifications of the $IIA$ string 
\cite{Wstd}, these singularities lead to non-perturbative gauge groups of the
corresponding $A$, $D$ or $E$ type.

Luckily there exist tables relating the Kodaira and the ADE types of
singularities.
For example, $I_0$ and $I_1$ correspond to smooth points of the surface,
whereas $I_n$ with $n\ge 1$ corresponds to $A_{n-1}$ and $II^*$
corresponds to $E_8$.
Far more information on K3 surfaces and Kodaira and ADE singularities is
given in \cite{Arev}.

Candelas and Font \cite{CF} discovered an extremely simple way of determining
the gauge group from the toric description of the variety:
Under favorable conditions the extended Dynkin diagrams of the gauge groups
can be read off from the polyhedron $\D^*$.
\begin{figure}[htb]
\epsfxsize=2.5in
\hfil\epsfbox{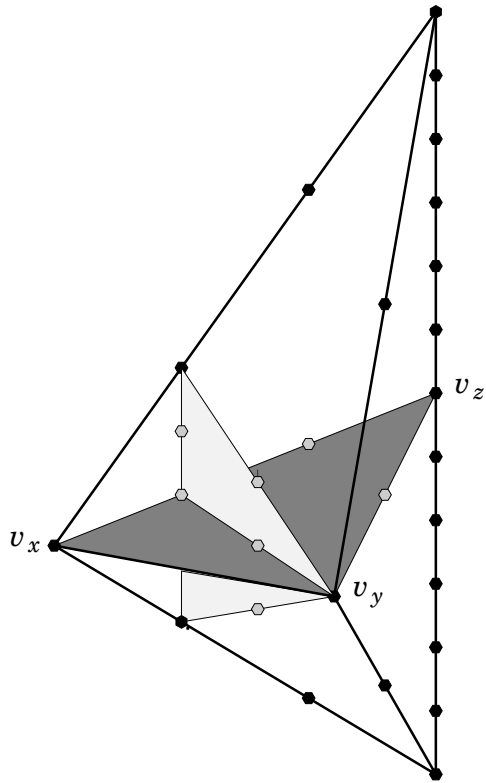}\hfil
\caption{The polyhedron $\D^*$ corresponding to the heterotic string with 
unbroken gauge symmetry}
\label{fig:poly}
\end{figure}
For example, fig. \ref{fig:poly} shows the polytope $\D^*$ dual to the 
polytope $\D$ determined by eq. (\ref{fibr}) with $a$ and $b$ as in 
(\ref{e8e8}).
The triangle lying `horizontally'
in $\D^*$ can be identified with the Weierstrass triangle.
Then the points above this triangle, together with the edges of $\D^*$ joining
them, form the extended Dynkin diagram of $E_8$.
As the points below the Weierstrass triangle form another $E_8$ extended
Dynkin diagram, we conclude that this configuration may be interpreted as the 
F-theory dual of the heterotic $E_8\times E_8$ string compactified on $T^2$
with no Wilson lines turned on.
It is interesting to note that the same polyhedron may also serve for
describing the F-theory dual of the $SO(32)$ string \cite{fst}:
If we slice it along the `vertical' triangle, we divide the polyhedron into
one part consisting of a single point, corresponding to a smooth fiber,
and a diagram that is just the extended Dynkin diagram of $SO(32)$.

The occurrence of Dynkin diagrams in toric polyhedra can be explained in the 
following way \cite{egs}:
The polytope $\D^*$ corresponds to the blown-up variety, so intersection
patterns of divisors correspond to structures of singularities.
Divisors in the K3 surface are intersections of divisors in the
ambient toric variety with the K3 surface, so intersections of two
divisors in the K3 surface correspond to triple intersections of
the form $D_1\cdot D_2\cdot K3$ in the toric variety.
The analysis of \cite{egs} shows that expressions of this form are
non-vanishing if and only if the points $v_1$ and $v_2$ corresponding to
$D_1$ and $D_2$, respectively, are either equal or joined by an edge of $\D^*$.
Then their intersection matrix is indeed the Cartan matrix of the 
corresponding Lie algebra. 

Compactifying F-theory to 6 dimensions, one usually assumes a double fibration
structure:
The Calabi-Yau threefold on which we compactify is supposed to be K3
fibered, with the K3 fiber being elliptically fibered. 
In terms of toric geometry this manifests itself as
\beq \D^*_{CY_3} \supset \D^*_{K3} \supset \D^*_{T^2}, \eeq
where each of the inclusions is such that the lower dimensional polyhedron
resides at a slice of the larger one through the origin.

Again the base of the elliptic fibration is the toric variety determined
by the fan that one obtains by projecting the fan for the ambient space
of the threefold along the directions of the elliptic fiber.
Gauge groups can be read off by looking at the preimages of rays in the
fan of the base.
These preimages, called `tops', again look like extended Dynkin diagrams of 
gauge groups.
For appropriately chosen polyhedra $\D^*$, these gauge groups may become
as large as $E_8\times(E_8\times F_4\times G_2^2\times A_1^2)^{16}$ for the
$E_8\times E_8$ string \cite{CPR3} and 
$SO(32)\times Sp(24)\times SO(80)\times Sp(48)\times SO(128)\times Sp(72)
   \times SO(176)\times Sp(40)\times Sp(52)\times SO(64)$
for the $SO(32)$ string \cite{AM,fst}.

{\it Acknowledgements:} 
This work is supported by the Austrian Research Funds
FWF (Schr"o\-din\-ger fellowship 
J1530-TPH), NSF grant PHY-9511632 and the Robert A. Welch Foundation.


\bye